\documentclass{vietnam}

\def\be{\begin{equation}}
\def\ee{\end{equation}}
\def\bea{\begin{eqnarray}}
\def\eea{\end{eqnarray}}

\begin{document}
\vspace*{4cm}
\hfill KA-TP-33-2013

\hfill SFB/CPP-13-73

\title{FULL NLO MASSIVE GAUGE BOSON\\
  PAIR PRODUCTION AT THE LHC}

\author{Julien Baglio$^{1}$ \footnote{speaker}, Le Duc Ninh$^{1,2}$ and
  Marcus M. Weber$^{3}$}

\address{$^{1}$ Institut f\"{u}r Theoretische Physik, Karlsruher
  Institut f\"{u}r Technologie,\\
  Wolfgang Gaede Strasse 1, Karlsruhe DE-76131, Germany\\
  $^{2}$ Institute of Physics, Vietnam Academy of Science and
  Technology,\\
  10 Dao Tan, Ba Dinh, Hanoi, Vietnam\\
$^{3}$ Max-Planck-Institut f\"{u}r Physik
(Werner-Heisenberg-Institut),\\ M\"{u}nchen D-80805, Germany}

\maketitle

\abstracts{Electroweak gauge boson pair production is a very
  important process at the LHC as it probes the non-abelian structure
  of electroweak interactions and is a background process for many
  searches. We present full next--to--leading order predictions for the
  production cross sections and distributions of on-shell massive
  gauge boson pair production in the Standard Model, including both
  QCD and electroweak corrections. The hierarchy between the $ZZ$,
  $WW$ and $WZ$ channels, observed in the transverse momentum distributions, 
  will be analyzed. We will also present a comparison with experimental data
  for the total cross sections including a study of the theoretical
  uncertainties.}

\section{Introduction}

Since the beginning of LHC operations in 2010, there have been
numerous gauge boson pair measurements at 7 and 8 TeV, in particular
looking for signs of new physics via anomalous
couplings~\cite{ATLAS-WW-7tev,CMS-WW-ZZ-8tev}. It
is indeed crucial to test the non-abelian structure of the electroweak (EW)
sector of the Standard Model (SM) as new physics effects could modify
this structure. When added to the fact that gauge boson pair
production is an important background in the search for the Higgs
boson, this triggers precise predictions on the theoretical side.

The QCD next--to--leading order (NLO) corrections have been known for
decades~\cite{WZ-QCD-NLO-1,WW-QCD-NLO,QCD-NLO-exclusive}. A full
next--to--next--to--leading order (NNLO) QCD calculation is not yet
available but some approximate results have been released in the past
few months, for example in the $WW$ production~\cite{Dawson:2013lya}. NLO
EW corrections, known for a long time in the high energy
approximation~\cite{EW-corrections-early1,EW-corrections-early2,EW-corrections-early3},
have been fully calculated only
recently~\cite{WW-EW-Kuhn,Kuhn-2,Baglio:2013toa} including
photon-quark induced processes~\cite{Baglio:2013toa}.

We present full NLO predictions for the total cross sections and
the differential distributions, including both QCD and EW effects. In
particular the hierarchy that is observed in the $p_T$ 
distributions between the $ZZ$, $WW$ and $WZ$ channels is explained
thanks to soft gauge boson approximation. A comparison with
experimental data including theoretical uncertainties is also
given. More details can be found in Ref.~\cite{Baglio:2013toa}.

\section{Overview of the calculation}

The well-known QCD NLO corrections to $q\bar{q}'\to VV'$ that were
calculated a while ago~\cite{WZ-QCD-NLO-1,WW-QCD-NLO,QCD-NLO-exclusive}
have been recalculated as well as the gluon fusion channel that is
formally a NNLO contribution. 
The NLO EW corrections include not only the virtual and real photon emission 
corrections to $q\bar{q}'$ channels but also the 
photon--quark induced channels, the latter not being considered in
Refs.~\cite{WW-EW-Kuhn,Kuhn-2}. The NLO corrections
to photon--photon initial state in the $WW$ channel were also
incorporated. We used the {\tt MRST2004QED} PDF set~\cite{PDF-QED} to
account for the photon PDF. The relevant EW parameters are
renormalized in the on--shell scheme.

In order to deal with infrared singularities we used dimensional
regularization and mass regularization schemes. The two calculations
are in excellent agreement. The Catani-Seymour dipole substraction
method~\cite{dipole-substraction} is used to combine the virtual 
and real corrections. We also cross-checked the results with the
phase-space slicing method~\cite{slicing-method}. We performed
independent calculations with the help of automated tools: 
the {\tt FeynArt/FormCalc}~\cite{FormCalc} suite to generate one-loop
amplitudes. The one-loop integrals are calculated with the in-house
library {\texttt{LoopInts}}, which agrees with the program
{\texttt{LoopTools}}~\cite{FormCalc,Form}. {\tt MadGraph} and {\tt
  HELAS} routines are also used to calculate tree-level
amplitudes. Further details about the calculation and the precise
definitions of the various contributions discussed  in the next
section can be found in Ref.~\cite{Baglio:2013toa}.

\section{Hierarchy of radiative corrections}

\begin{figure}
\begin{minipage}{0.32\linewidth}
\centerline{\includegraphics[width=0.9\linewidth]{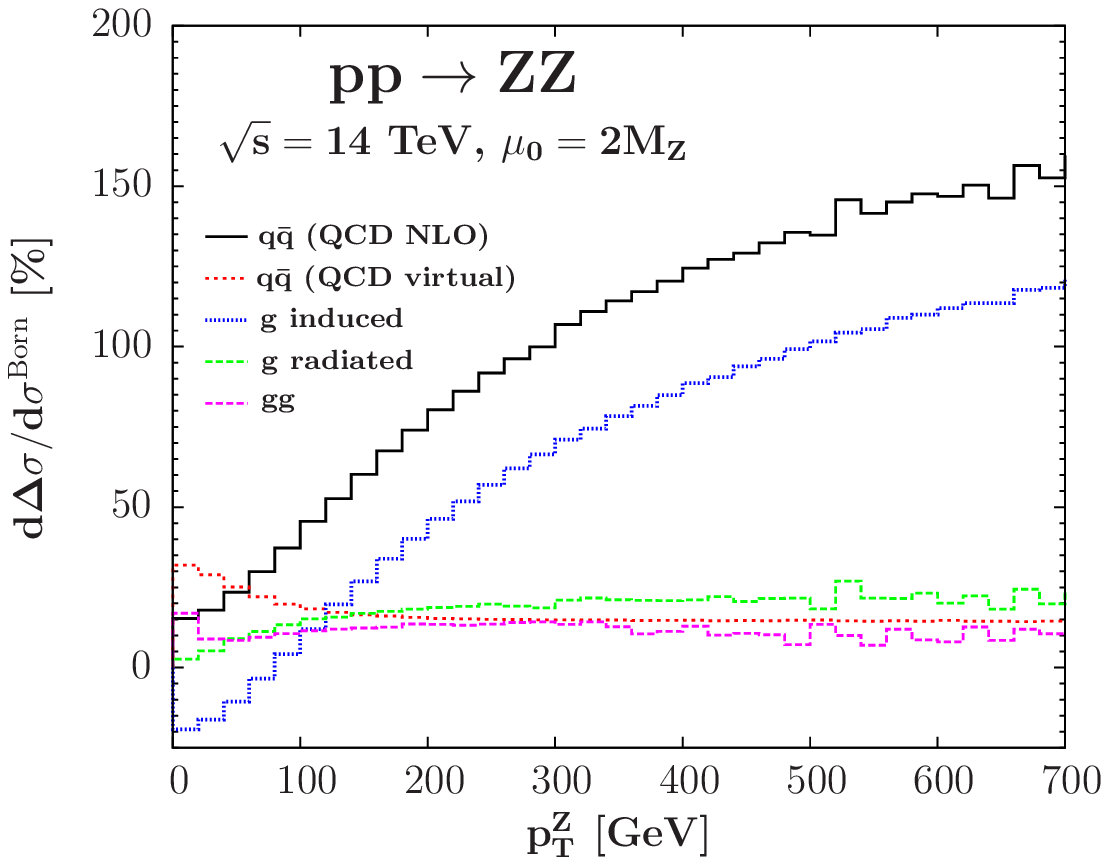}}
\end{minipage}
\hfill
\begin{minipage}{0.32\linewidth}
\centerline{\includegraphics[width=0.9\linewidth]{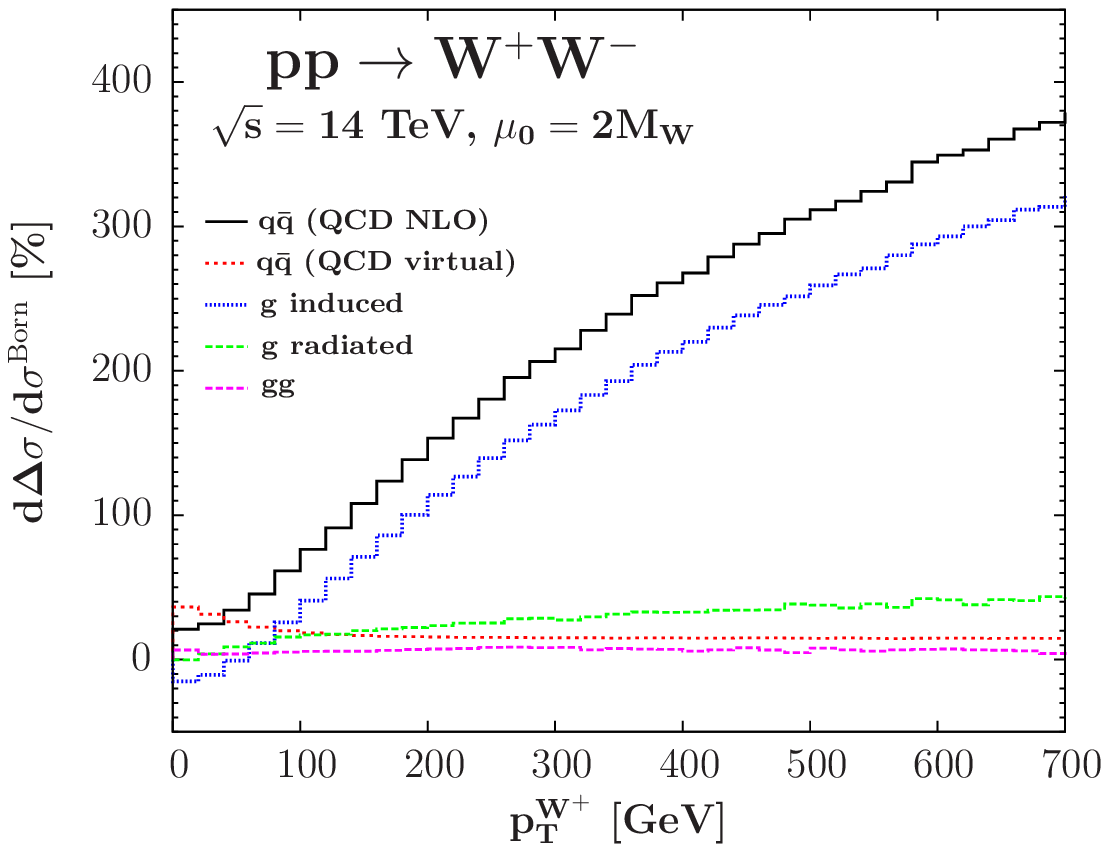}}
\end{minipage}
\hfill
\begin{minipage}{0.32\linewidth}
\centerline{\includegraphics[width=0.9\linewidth]{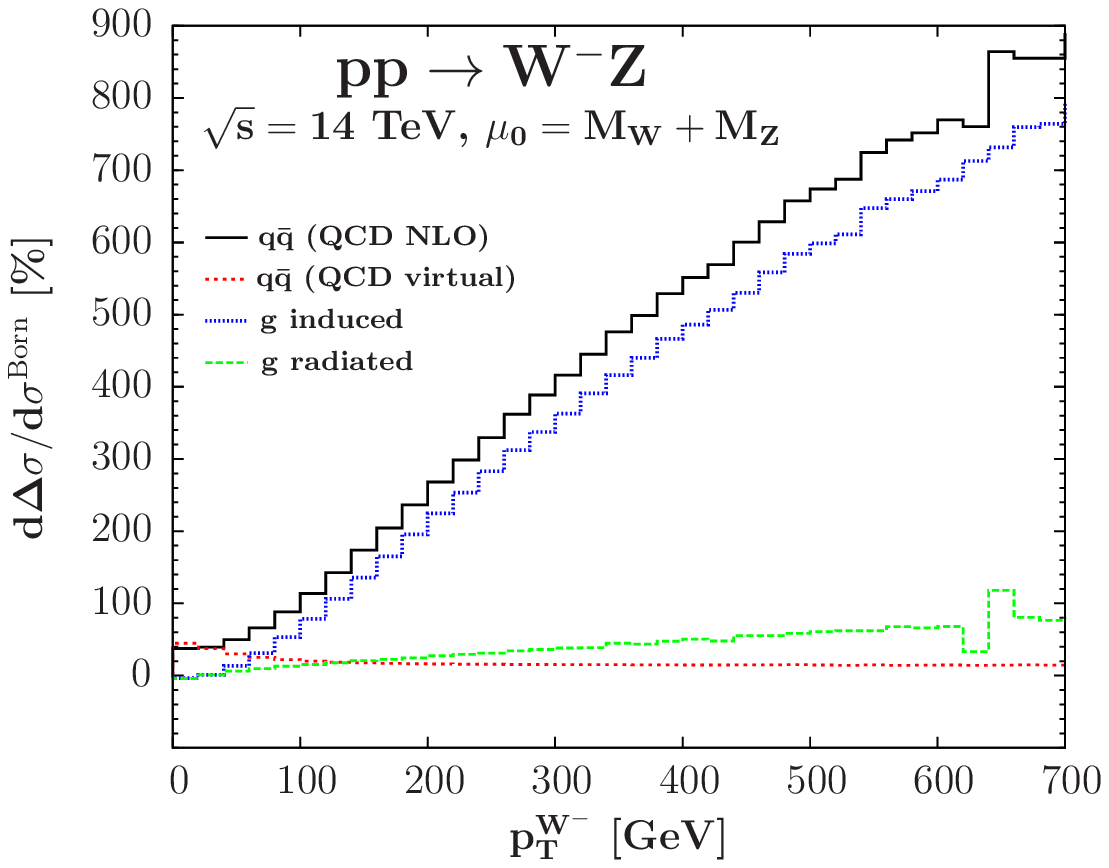}}
\end{minipage}
\caption{$Z$ (left), $W^+$ (middle) and $W^-$ (right) transverse
  momentum distributions (in GeV) of the NLO QCD corrections (in
  $\%$) in $\sigma(pp\to ZZ,\, WW,\, W^-Z)$, respectively.}
\label{fig:distribution_QCD}
\end{figure}

We present some selected results for the differential
distributions at the LHC at 14 TeV, using the {\tt MRST2004QED} PDF
set and $\alpha_s(M_Z^2)=0.1190$. The factorization and
renormalization scales  are both equal to $M_V + M_{V^\prime}$.
We apply no cuts at the level of the on-shell 
$W^\pm$ and $Z$, since these will decay. It can be seen in
Fig.~\ref{fig:distribution_QCD} that the QCD corrections are driven by
the gluon-quark induced processes (dotted blue lines) with a large
correction at high $p_T$ driven by leading-logarithmic terms
proportional to $\alpha_s \log^2(M_V^2/p_T^2)$. This is explained by
the large gluon PDF and soft gauge boson emission and has been noticed 
for quite a while~\cite{WW-QCD-NLO,QCD-NLO-exclusive} but the
hierarchy $\displaystyle \delta^{\rm QCD}_{ZZ} \simeq\frac13
\delta^{\rm QCD}_{WW} \simeq \frac16 \delta^{\rm QCD}_{W^- Z}$ with
$\delta_{ZZ}^{\rm QCD} \simeq 120\%$, clearly visible on
Fig.~\ref{fig:distribution_QCD}, was not well understood.

\begin{figure}
\begin{minipage}{0.32\linewidth}
\centerline{\includegraphics[width=0.9\linewidth]{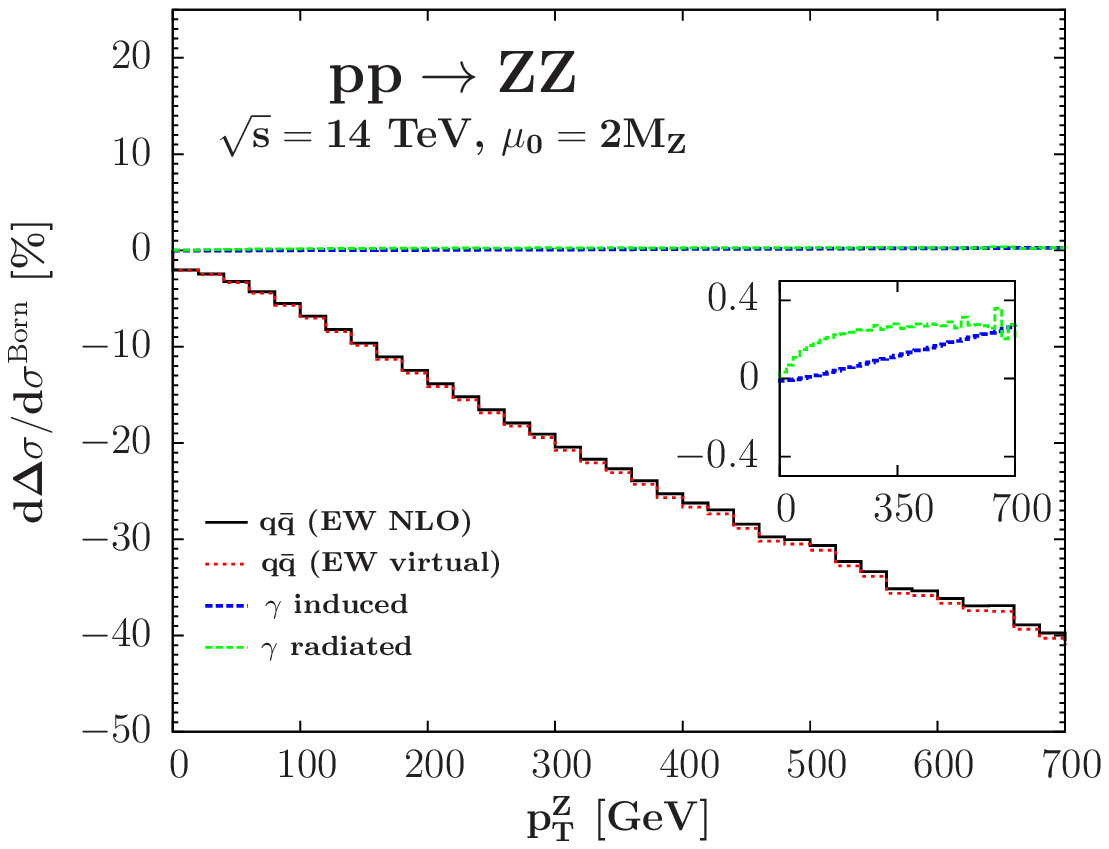}}
\end{minipage}
\hfill
\begin{minipage}{0.32\linewidth}
\centerline{\includegraphics[width=0.9\linewidth]{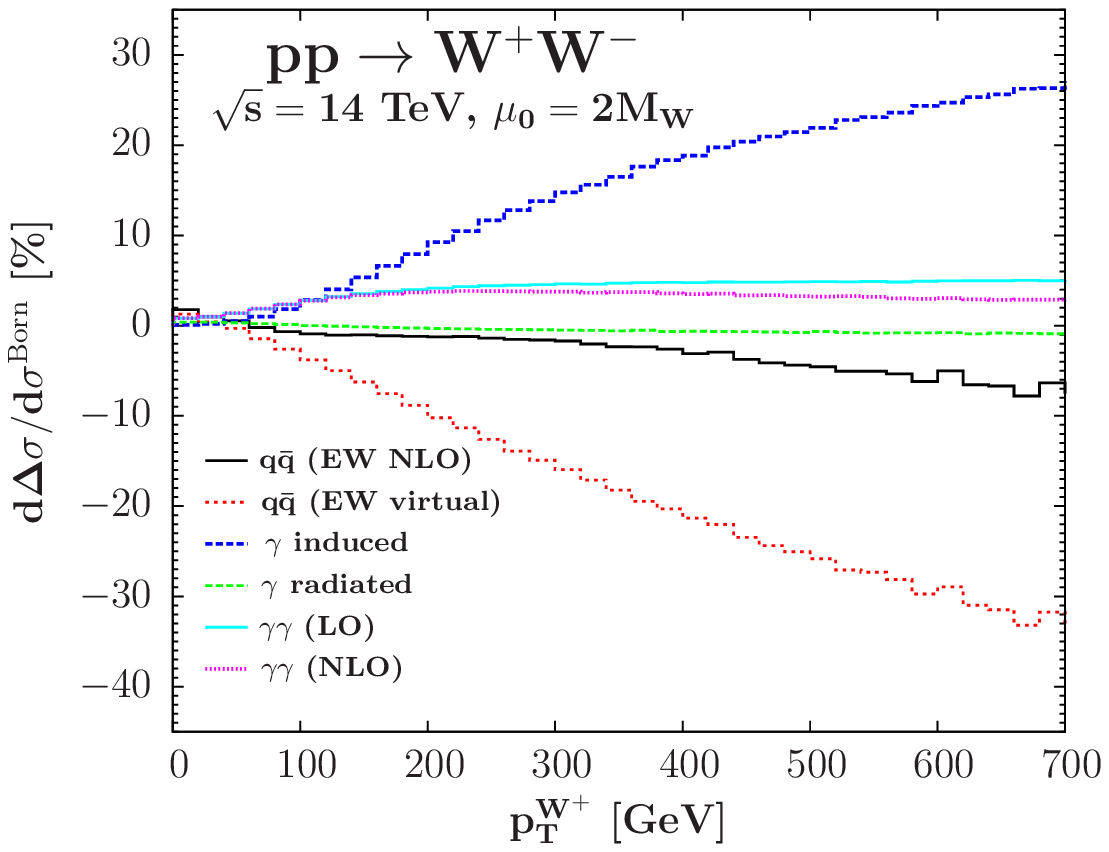}}
\end{minipage}
\hfill
\begin{minipage}{0.32\linewidth}
\centerline{\includegraphics[width=0.9\linewidth]{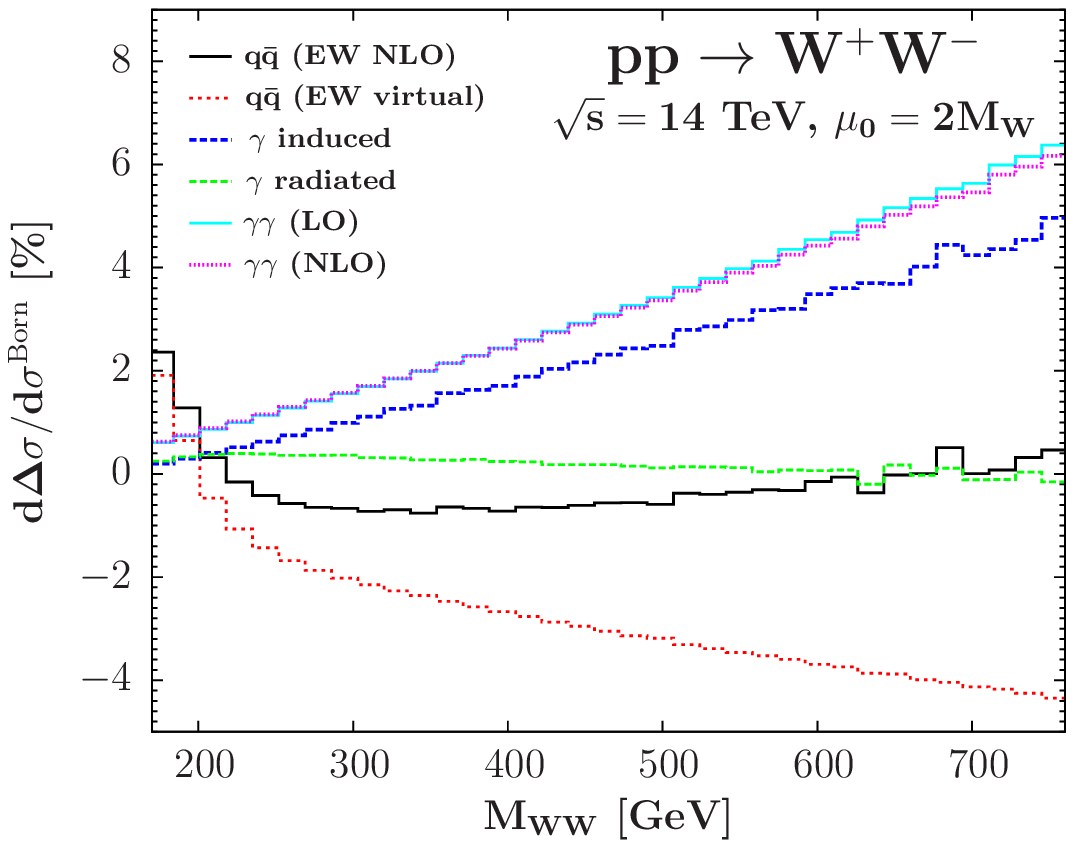}}
\end{minipage}
\caption{$Z$ (left) and $W^+$ (middle) transverse momentum
  as well as $M_{WW}$ invariant mass (right) distributions (in
  GeV) of the NLO EW corrections (in $\%$), in $\sigma(pp\to ZZ,\,
  WW,\, W^-Z)$, respectively.}
\label{fig:distribution_EW}
\end{figure}

The same hierarchy is also observed in the EW corrections as displayed
in Fig.~\ref{fig:distribution_EW} (left-handed and middle figures). This
hierarchy is much more pronounced than for the QCD case, with $\displaystyle
\delta^{\rm EW}_{ZZ} \simeq \frac{1}{90} \delta^{\rm EW}_{WW} \simeq
\frac{1}{190} \delta^{\rm EW}_{W^- Z}$ and $\delta_{ZZ}^{\rm EW}
\simeq 0.3\%$. The virtual Sudakov effect in the $q\bar{q}' \to V V'$ 
is clearly visible (dashed red lines) and has been
also discussed in Ref.~\cite{WW-EW-Kuhn,Kuhn-2}.

The hierarchies between the $ZZ$, $WW$ and $WZ$ channels in both QCD and
EW radiative corrections share some common features. They are 
effects of the dominant double-logarithmic terms in the gluon-quark and
photon-quark induced processes, in which the non-abelian structure of
the theory, the different couplings strengths and PDF effects play a 
role. The analytical approximations using soft gauge boson emission
from a quark-gauge boson final state, presented in
Ref.~\cite{Baglio:2013toa}, reproduce this hierarchy even if they are
off by a factor of two at $p_T\simeq 700$ GeV. It has been checked
that they coincide with the full result at much higher $p_T$, thereby 
validating the approximation. In the case of the EW corrections, the
$\gamma q$ processes are further enhanced by a $t$--channel
massive gauge boson exchange, explaining the huge enhancement of the
EW corrections in the $WW$ and $WZ$ channels compared to the $ZZ$
channel. It compensates or even overcompensates the virtual Sudakov
effect in the $WW$ and $WZ$ channels, making the photon-quark induced
processes absolutely necessary for the full NLO EW calculation. The
right-handed side of Fig.~\ref{fig:distribution_EW} shows the importance
of the diphoton subprocess in the $WW$ invariant mass distribution
where it is the leading EW effect.

\section{Total cross sections and comparison with experimental data}

We have calculated the total cross sections fully at NLO and compared
with the most up-to-date ATLAS and CMS results from HEP-EPS 2013
Conference. This is an update of our previous
results~\cite{Baglio:2013toa}. In order to account for the EW
corrections using modern PDF sets such as the {\tt MSTW}
set~\cite{PDF-MSTW}, we rescaled our NLO QCD predictions calculated
with modern sets by a factor $\delta^{\rm EW}$ calculated with {\tt
  MRST2004QED} including the photon PDF:
$\displaystyle \delta^{\rm EW} = \sigma^{\rm NLO}_{\rm QCD+EW}/\sigma^{\rm NLO}_{\rm QCD}$. 
Recently the NNPDF
Collaboration has released a new set including also a photon
PDF~\cite{NNPDF-QED} and we have checked at the level of the total
cross section that the ratio $\delta^{\rm EW}$ does not change
significantly by trading {\tt MRST2004QED} with {\tt NNPDF2.3QED}.

\begin{figure}
\begin{minipage}{0.36\linewidth}
\centerline{\includegraphics[width=0.9\linewidth]{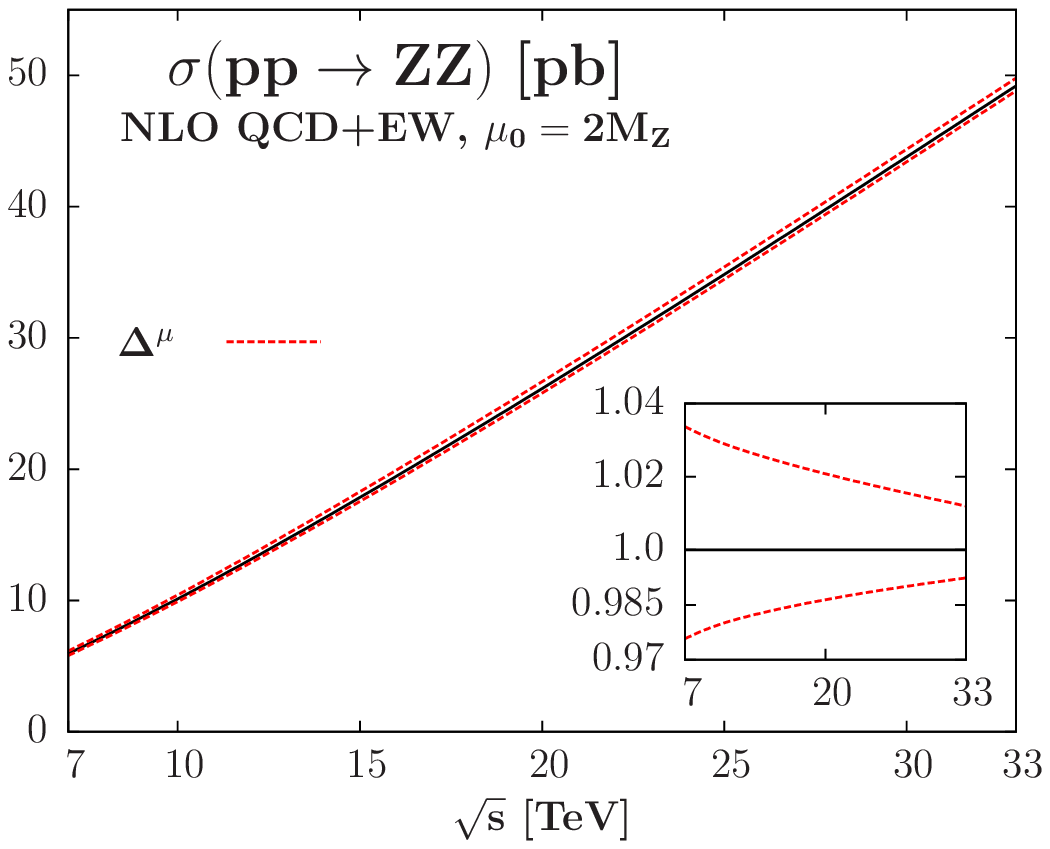}}
\end{minipage}
\hfill
\begin{minipage}{0.36\linewidth}
\centerline{\includegraphics[width=0.9\linewidth]{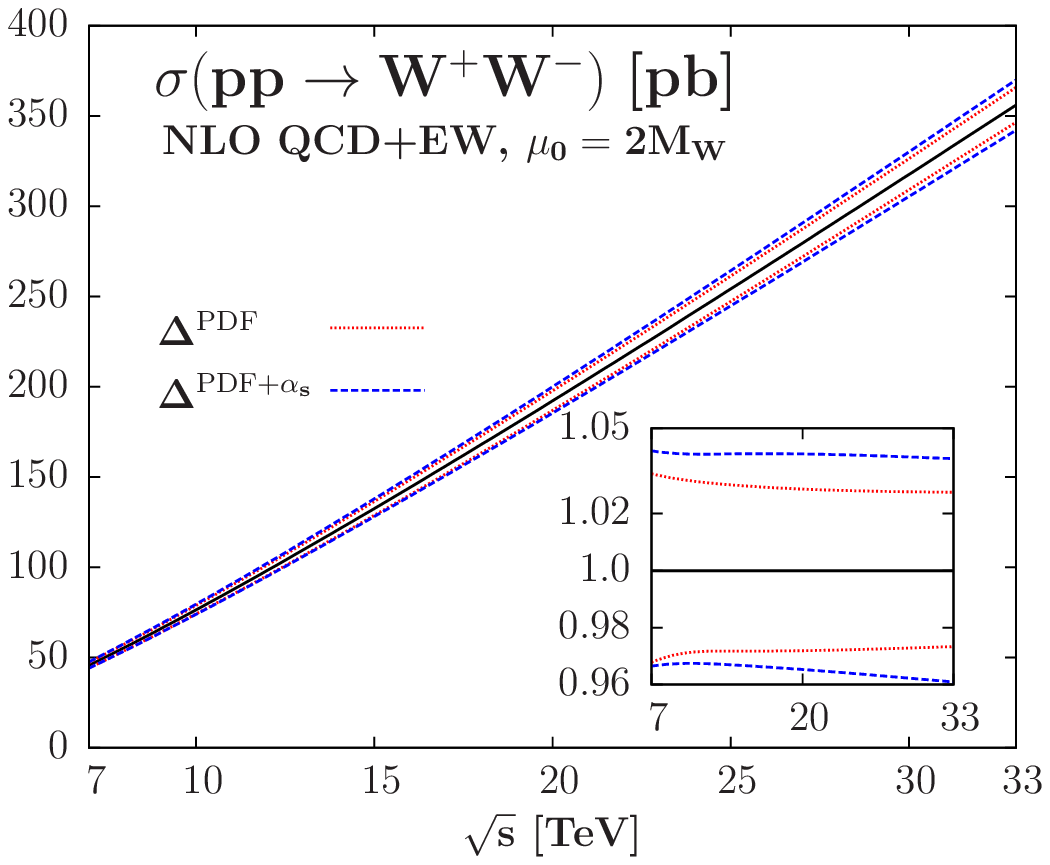}}
\end{minipage}
\hfill
\caption{The scale uncertainty in $\sigma(pp\to ZZ)$ at the LHC (left)
  and the PDF/PDF+$\alpha_s$ uncertainty in $\sigma(pp\to WW$ (right),
  in pb, as a function of the center-of-mass energy (in TeV). In the
  inserts deviations from the central predictions are shown.}
\label{fig:total_xs_pdf_scale_error}
\end{figure}

A detailed study of the theoretical uncertainties affecting the
predictions has been performed. We calculated the scale uncertainty
with the factorization and renormalization scales varied in the interval
$\frac12 \mu_0 \leq \mu_R=\mu_F \leq 2 \mu_0$ where the central scale
is $\mu_0=M_V + M_{V'}$. We used the {\tt MSTW2008} $90\%$CL set
to calculate the correlated PDF$+\alpha_s$ uncertainty. The parametric
uncertainties coming from the experimental errors on $M_W$ and $M_Z$
are negligible. The results are presented in
Fig.~\ref{fig:total_xs_pdf_scale_error} and are similar in the three
different channels. We obtain $\delta^{\rm EW} \approx 0.97, 1.00, 1.01$
for the $ZZ,WZ,WW$ channels respectively. The scale uncertainty
amounts to $\simeq +3\% / -2\%$ at 7 TeV, two times less at 33
TeV. The PDF+$\alpha_s$ uncertainty is of the order of $\pm 4\%$.

\begin{figure}
\begin{minipage}{0.32\linewidth}
\centerline{\includegraphics[width=0.9\linewidth]{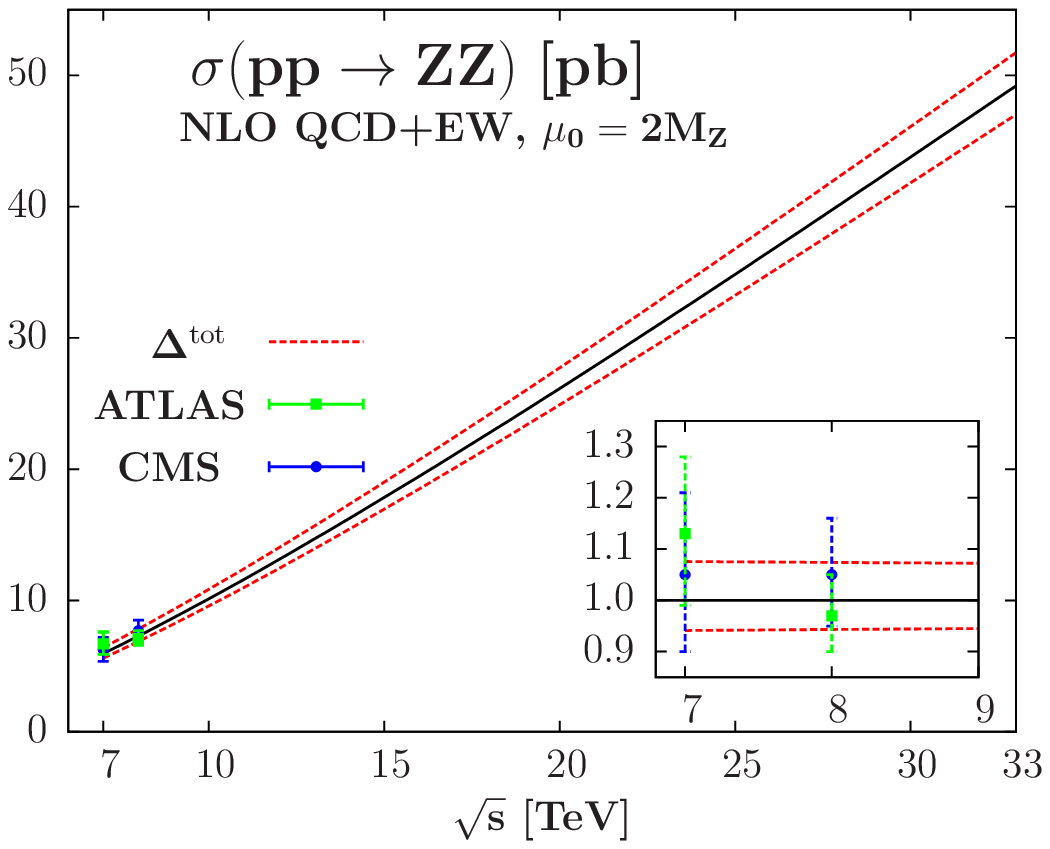}}
\end{minipage}
\hfill
\begin{minipage}{0.32\linewidth}
\centerline{\includegraphics[width=0.9\linewidth]{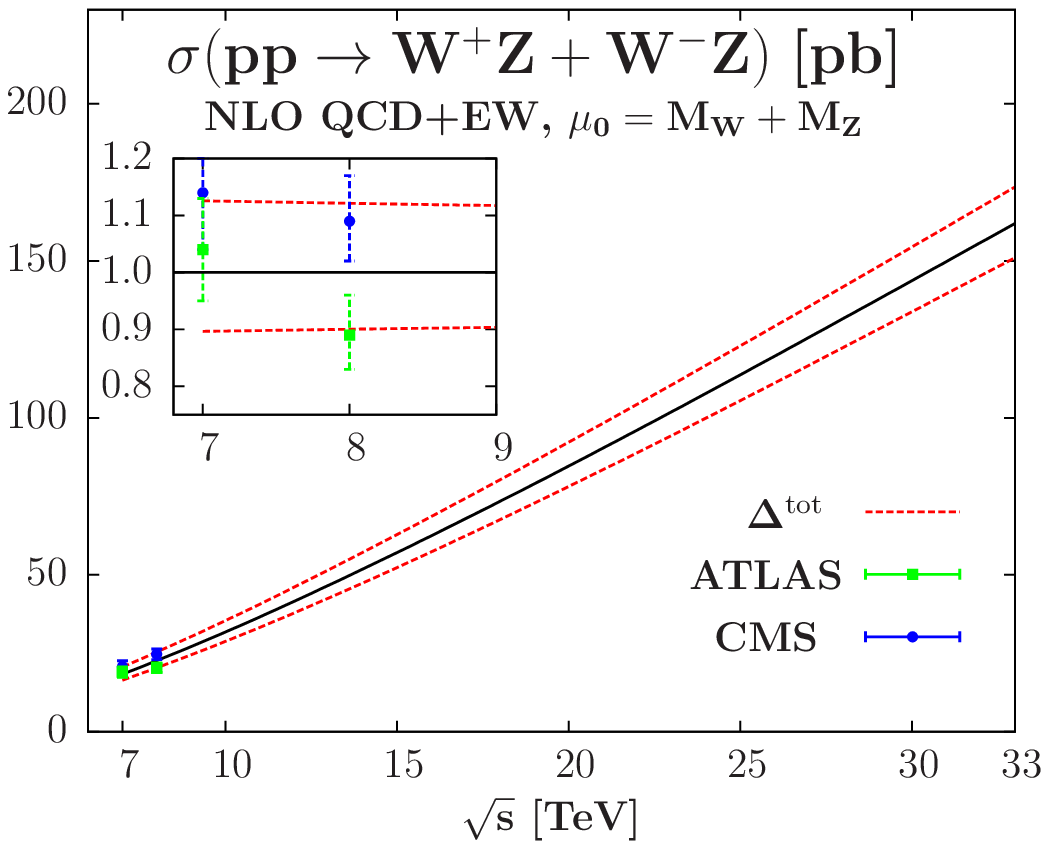}}
\end{minipage}
\hfill
\begin{minipage}{0.32\linewidth}
\centerline{\includegraphics[width=0.9\linewidth]{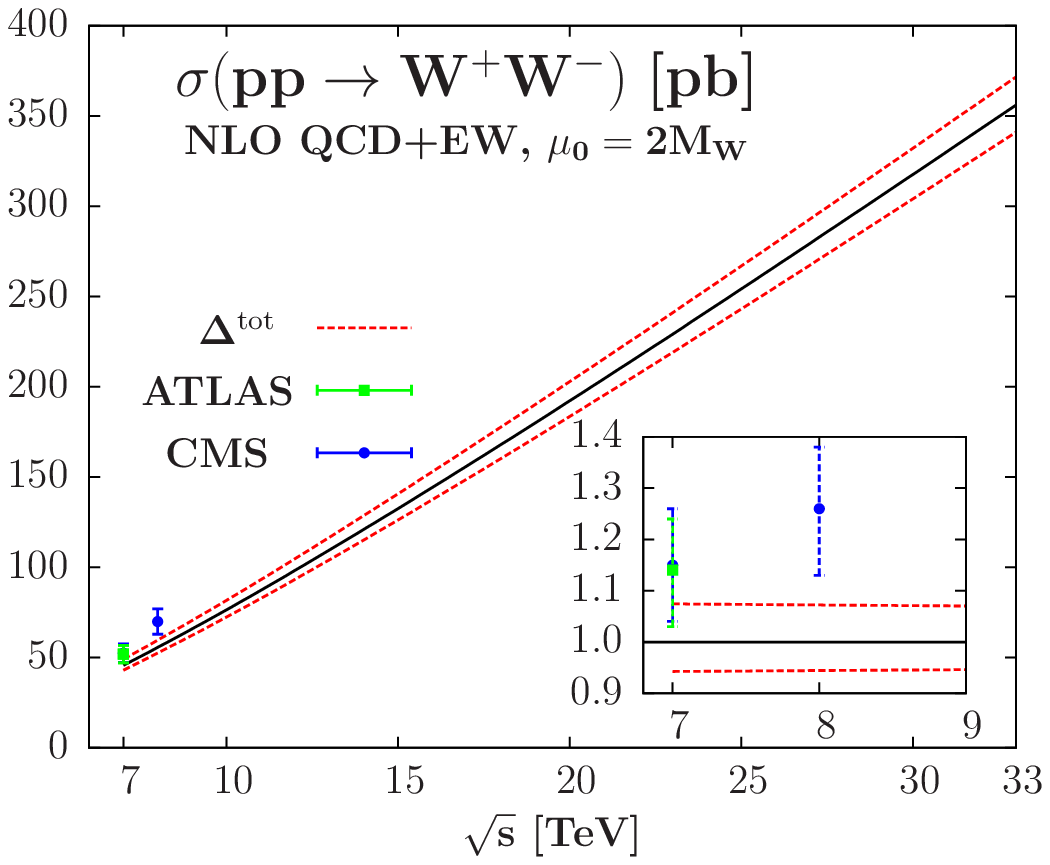}}
\end{minipage}
\caption{The NLO QCD+EW total cross section (in pb) of the processes
  $pp \to ZZ$ (left), $pp\to W^+Z + W^- Z$ (middle) and $pp\to WW$
  (right) at the LHC as a function of the center-of-mass energy (in
  TeV) including the total theoretical uncertainty. The insert shows
  the relative deviation from the central cross sections, and the
  experimental data points are also displayed on the main figures.}
\label{fig:total_xs_total_error}
\end{figure}

We have combined the scale and PDF+$\alpha_s$ into the overall
theoretical uncertainty of the total cross section and compared with
experimental data, as displayed in
Fig.~\ref{fig:total_xs_total_error}. We found a total uncertainty
$\sim +7\% / -6\%$ at 7 and 8 TeV, slightly less at 33 TeV, in all
three channels. The comparison with ATLAS and CMS results is good, in
particular in the $ZZ$ and $WZ$ channels. In the $WW$ channel, there
is a $1\sigma$ excess at 7 TeV and a $1.8\sigma$ excess at 8 TeV. As
estimated in Ref.~\cite{Baglio:2013toa} and confirmed by the results
in Ref.~\cite{Dawson:2013lya} a full NNLO calculation is not expected to 
account for this deviation.

\section*{Acknowledgments}

J.B. would like to thank the organizers for the very nice atmosphere
of the conference. This work is supported by the Deutsche
Forschungsgemeinschaft via the Sonderforschungsbereich/Transregio
SFB/TR-9 Computational Particle Physics.

\bibliography{windows_on_the_universe_vv_production}

\end{document}